\index{}
\documentclass[longauth,structabstract]{aa}

\usepackage{graphicx}
\usepackage{ulem}
\usepackage{xcolor}
\usepackage{txfonts}

\usepackage{natbib}
\usepackage{multirow}

\newcommand{\lat}{\textit{Fermi}--LAT }

\begin{document}

\title{MAGIC observations and multifrequency properties of the flat spectrum radio quasar \object{3C~279} in 2011}

\author{
 J.~Aleksi\'c\inst{1} \and
S.~Ansoldi\inst{2} \and
L.~A.~Antonelli\inst{3} \and
P.~Antoranz\inst{4} \and
A.~Babic\inst{5} \and
P.~Bangale\inst{6} \and
U.~Barres de Almeida\inst{6} \and
J.~A.~Barrio\inst{7} \and
J.~Becerra Gonz\'alez\inst{8} \and
W.~Bednarek\inst{9}
% ...
}
 \institute{ 
IFAE, Edifici Cn., Campus UAB, E-08193 Bellaterra, Spain
\and Universit\`a di Udine, and INFN Trieste, I-33100 Udine, Italy
\and INAF National Institute for Astrophysics, I-00136 Rome, Italy
\and Universit\`a  di Siena, and INFN Pisa, I-53100 Siena, Italy
\and Croatian MAGIC Consortium, Rudjer Boskovic Institute, University of Rijeka and University of Split, HR-10000 Zagreb, Croatia
\and Max-Planck-Institut f\"ur Physik, D-80805 M\"unchen, Germany
\and Universidad Complutense, E-28040 Madrid, Spain
\and Inst. de Astrof\'isica de Canarias, E-38200 La Laguna, Tenerife, Spain
\and University of \L\'od\'z, PL-90236 Lodz, Poland
% ...
}

 
  \abstract
   {}
   {We study the multifrequency emission and spectral properties of the quasar \object{3C~279} aimed at identifying the radiation  processes taking place in the source.}
   {We observed \object{3C~279} in very-high-energy (VHE, E $>100\,$GeV) $\gamma$-rays, with the MAGIC telescopes during 2011, for the first time in stereoscopic mode. We combined these measurements with observations at other energy bands: 
in high-energy (HE, E $>100\,$MeV) $\gamma$-rays from \textit{Fermi}--LAT; in X-rays from RXTE; in the optical from the KVA telescope; and in the radio at 43\,GHz, 37\,GHz, and 15\,GHz from the VLBA, Mets\"ahovi, and OVRO radio telescopes - along with optical polarisation measurements from the KVA and Liverpool telescopes. We examined the corresponding light curves and broadband spectral energy distribution and we compared the multifrequency properties of \object{3C~279} at the epoch of the MAGIC observations with those inferred from historical observations.}
    {During the MAGIC observations (2011 February 8 to April 11) \object{3C~279} was in a low state in optical, X-ray, and $\gamma$-rays. The MAGIC observations did not yield a significant detection. The derived upper limits are in agreement with the extrapolation of the HE $\gamma$-ray spectrum, corrected for EBL absorption, from \textit{Fermi}--LAT. The second part of the MAGIC observations in 2011 was triggered by a high-activity state in the optical and $\gamma$-ray bands. During the optical outburst the optical electric vector position angle (EVPA) showed a rotation of $\sim180^\circ$. Unlike previous cases, there was no simultaneous rotation of the 43\,GHz radio polarisation angle. No VHE $\gamma$-rays were detected by MAGIC, and the derived upper limits suggest the presence of a spectral break or curvature between the \lat and MAGIC bands. The combined upper limits are the strongest derived to date for the source at VHE and below the level of the previously detected flux by a factor of $\sim2$. Radiation models that include synchrotron and inverse Compton emissions match the optical to $\gamma$-ray data, assuming an emission component inside the broad line region with size $R=1.1\times 10^{16}\,$cm and magnetic field $B=1.45\,$G responsible for the high-energy emission, and another one outside the broad line region and the infrared torus ($R=1.5\times 10^{17}\,$cm and $B=0.8\,$G) causing the optical and low-energy emission.
    We also study the optical polarisation in detail and interpret it with a bent trajectory model.}
{}


\titlerunning{MAGIC observations and multifrequency properties of the FSRQ \object{3C~279} in 2011}

\maketitle

\section{Introduction}

Blazars, active galactic nuclei (AGNs) with the relativistic jets oriented at small angles with respect to the line of sight \citep{Urry1995}, constitute the most numerous class of very-high-energy (VHE, E $> 100$ GeV) $\gamma$-ray emitters. Nowadays, we count around fifty\footnote{http://tevcat.uchicago.edu/} members of this class, which is further 
divided into BL Lac objects (BL Lacs) and flat spectrum radio quasars (FSRQs). In the VHE range only three $\gamma$-ray sources belonging to this latter class have been detected, i.e. \object{3C~279} \citep{Albert2008Science}, PKS~1222+216 \citep{Aleksic2011b}, and PKS~1510$-$089 \citep{hess2013,Aleksic_2014}. 

All blazars are highly variable, emitting nonthermal radiation spanning more than ten orders of magnitude in energy, and they show distinct features, in particular in the optical spectrum. BL Lacs are characterised by a continuous spectrum with weak or no emission lines in the optical regime while FSRQs show broad emission lines. Consequently, blazars are classified as BL Lacs or FSRQs according to the width  of the strongest optical emission line, which is $<5\AA{}$ in BL Lacs \citep{Urry1995}. 
The presence of emission lines has several implications. In combination with the often observed big blue bump in the optical-UV region from the accretion disc, the presence of gas and low-energy radiation around these sources is suggested. This has further implications for emission models; the VHE emission may be absorbed by internal optical and UV radiation coming from the accretion disc or from the broad line region (BLR). Therefore, it is reasonable to assume the presence of a population of low-energy photons coming from either one of these regions or from both of them, which contributes to the overall observed emission. Furthermore, pronounced emission lines allow for a good measurement of the redshift, which is usually precisely determined for FSRQs while for BL Lacs it is often unknown or limited to a range of values. The traditional classification of blazars into BL Lacs and FSRQs, outlined above, has recently been called into question \citep{Giommi2012}. 

The spectral energy distribution (SED) of blazars has two broad peaks, the first between mm wavelengths and soft X-ray wavelengths, the second in the MeV/GeV band \citep{Ghisellini2008}. Typically, FSRQs have lower peak energies and higher bolometric luminosity than BL Lacs. In addition, their high-energy peak is the more prominent \citep[e.g. see Compton dominance distributions in][]{Giommi2012observations} . Various scenarios have been proposed to explain the emission of blazars. The low-energy peak is believed to be associated with synchrotron radiation from relativistic electrons, while for the high-energy peak there is no general agreement, and different models are used for particular sources. For most BL Lacs, the second peak is explained as Compton up-scattering of the low-energy photons. Target photons can be the low-energy photons of the synchrotron emission \citep[SSC; synchrotron self-Compton,][]{Band1985} or, in the case of External Compton models \citep[EC; e.g.][]{Hartman2001, Boettcher2013}, the seed photons are provided by the accretion disc, BLR clouds, and dusty torus.
The case of FSRQs is different. Initially, at the time of early $\gamma$-ray observations, synchrotron self-Compton models \citep{Maraschi1992, Maraschi1994} and hadronic self-Compton models \citep{Mannheim1992} were applied to FSRQs. Later, it was found that the short variability timescales observed seemed to favour leptonic emission, since the acceleration timescale for protons is much longer. If, however, the variability is governed by  the dynamical timescale, hadronic emission is a viable explanation, provided that the proton energies are high enough to guarantee a high radiative efficiency. External Compton models \citep[e.g.][]{Hartman2001}, models with several emission zones \citep[e.g.][]{Tavecchio2011} and further extensions of hadronic models have been proposed.
\smallskip

The source \object{3C~279} was the first $\gamma$-ray quasar discovered with the \textit{Compton Gamma-Ray Observatory} \citep{Hartman1992} and is the first member of the class of FSRQs detected as a VHE $\gamma$-ray emitter \citep{Albert2008Science}. In addition, with a redshift of 0.536, it is among the most distant VHE $\gamma$-ray extragalactic sources detected so far. 
VHE $\gamma$-rays interact with low-energy photons of the extragalactic background light (EBL) via pair-production, making the source visibility in this energy range dependent on its distance. The discovery of \object{3C~279} as a $\gamma$-ray source stimulated debate about the models of EBL available at that time, implying a lower level of EBL than thought. Furthermore, the discovery of this source had interesting implications for emission models. Simple one-zone SSC models were not able to explain the observed emission requiring the development of more complicated scenarios and hadronic models \citep{Boettcher2009,Aleksic2011a}. In addition, different models need to be considered for different activity states. \citet{Boettcher2013} could not fit the SED of a low-activity state of \object{3C~279} with a hadronic model while \citet{Boettcher2009} provides satisfactory hadronic fits of a flaring state.

\smallskip 

Recently several papers have been published reporting large rotations ($> 180^\circ$) of the optical electric vector position angle (EVPA) in high-energy (HE, $100\,$MeV $<$ E $<100\,$GeV) and VHE $\gamma$-ray emitting blazars: \object{3C~279} \citep{Larionov2008}, BL Lacertae \citep{Marscher2008}, and PKS 1510-089 \citep{Marscher2010}. In almost every case the rotations appear in connection with $\gamma$-ray flares and high-activity states of the sources. These long, coherent rotation events have been interpreted as the signature of a global field topology or the geometry of the jet, which are traced by a moving emission feature. 
For the case of \object{3C~279}, two such rotation events have been detected. The first one \citep{Larionov2008} was associated with the $\gamma$-ray flare detected by MAGIC \citep{Aleksic2011a}, whereas the second \citep{Abdo2010Nature} was observed in conjunction with a HE $\gamma$-ray flare detected by the \textit{Fermi} Large Area Telescope (LAT) and interpreted as the signature of a bend in the jet a few parsecs downstream from the AGN core. In this work, an EVPA rotation that happened around MJD 55720 is reported. This event is therefore the third episode of large EVPA rotation detected for \object{3C~279}. While the rotation events seem to be rather common in $\gamma$-ray emitting blazars during the $\gamma$-ray flares, the connection between them and the HE and VHE emission in blazars is still under discussion. 
Historically, variations of the circularly polarised flux in the optical band have also been measured \citep{Wagner2001} supporting the idea that flux enhancements can go along with magnetic field structure.

\section{MAGIC observations and data analysis}

\begin{table*}
\centering
\caption{Results of 2011 MAGIC observations. For both the individual observation periods and for the entire 2011 data set, the observation time in hours, the excess and background events, and the significance calculated with Eq. 17 of \citet{LiMa1983}, are reported.}
\label{magicr}
\begin{tabular}{c c c c c}
  \hline \hline

  Observation period & Observation time [h] & Excess events [counts] & Background events [counts] & significance \\ \hline
2011 Feb - Apr & 11.6 & 34 $\pm$  82 & 3354  $\pm$ 58 & 0.4 $\sigma$  \\ 
2011 Jun & 6.2 & 46 $\pm$  60 & 1790  $\pm$ 42 & 0.8 $\sigma$ \\ 
all 2011 data & 17.9 & 80 $\pm$ 102 & 5144 $\pm$ 72 & 0.8 $\sigma$\\ 
\hline
\end{tabular}

\end{table*}

  \begin{figure}
 \includegraphics[width=88mm]{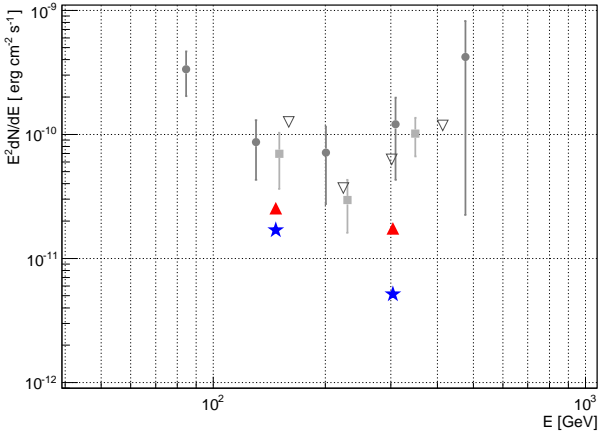}
\caption{Differential upper limits calculated from MAGIC observations from the two individual observations periods in 2011 (blue stars for February-April upper limits, red filled triangles for June upper limits). Previous MAGIC-I observations are also shown \citep{Aleksic2011a}: the 2006 discovery (grey circles), 2007 detection (grey squares), and the upper limits derived from the 2009 observations (grey open down-pointing triangles). All observations are corrected for EBL absorption using \citet{Dominguez2011}.}
\label{vhe_sed}  
  \end{figure}

Very-high-energy $\gamma$-ray observations were performed with the MAGIC telescopes, a system of two 17\,m diameter imaging Cherenkov telescopes located on the Canary Island of La Palma, at the observatory of the Roque de Los Muchachos (28.8$^\circ\,$ N, 17.8$^\circ\,$ W at 2200\,m a.s.l). The stereoscopic system provided an energy threshold of 50\,GeV and a sensitivity of $(0.76 \pm 0.03)$\,\% of the Crab Nebula flux, for 50 hours of effective observation time in the medium energy range above 290\,GeV \citep[for details see][]{Aleksic2012}. 
Because of the limited field of view ($\sim3.5^\circ$) of the MAGIC telescopes, we did not operate in surveying mode, but we tracked selected sources. One of the most successful techniques for discovering new sources or detecting flaring states is a target of opportunity (ToO) program triggered by an alert of a high-activity state in other wavebands.

The data analysis was performed using MARS \citep{Moralejo2009}, the standard MAGIC analysis framework with adaptations for stereoscopic observations \citep{Lombardi2011}. Based on the timing information, an image cleaning was performed with absolute cleaning levels of 6 photoelectrons (so-called core pixels) and 3 photoelectrons (boundary pixels) for the MAGIC-I telescope and 9 photoelectrons and 4.5 photoelectrons for the MAGIC-II telescope  \citep{Aliu2009}. The shower arrival direction is reconstructed using a random forest regression method \citep{Aleksic2010}, extended with stereoscopic information such as the height of the shower maximum and the impact distance of the shower on the ground \citep{Lombardi2011}. In order to distinguish $\gamma$-like events from hadron events, a random forest method is applied \citep{Albert2008b}. In the stereoscopic analysis image parameters of both telescopes are used, following the prescription of \citet{Hillas1985}, as well as the shower impact point and the shower height maximum. We additionally reject events whose reconstructed source position differs by more than $(0.05^\circ)^2$ in each telescope. A detailed description of the stereoscopic MAGIC analysis can be found in \citet{Aleksic2012}. 

The source \object{3C~279} was observed in 2011 as part of two different campaigns. Initially, it was observed for about 20 hours, during 14 nights from February 8 to April 11 for regular monitoring. In June, high-activity states in the optical and \lat energy ranges triggered ToO observations. The source was observed for a total of about 10 hours on 7 nights (from 2011 June 1 to 2011 June 7). Hereafter, the February to April observations and the June observations refer to the periods of MAGIC observations. After a quality selection based on the event rate, excluding runs with bad weather and technical problems, the final data sample amounts to 20.58 hours. The effective time of these observations, corrected for the dead time of the trigger and readout systems, is 17.85 hours. Part of the data was taken under moderate moonlight 
and twilight conditions, and these were analysed together with those taken during dark nights \citep{Britzger2009}. The source was observed at  high zenith angles, between 35$^\circ$ and 45$^\circ$. All data were taken in the false-source tracking (wobble) mode \citep{Fomin1994}, in which the telescope pointing was alternated every 20 minutes between two sky positions at 0.4$^\circ$ offset from the source, with a rotation angle of 180$^{\circ}$. This observation mode allows us to take On and Off data simultaneously. The background is estimated from the anti-source, a region located opposite to the source position.

For all 2011 observations, above 125\,GeV the distribution of the squared angular distance between the pointed position and the reconstructed position in the MAGIC data indicates an excess of 80 $\pm$ 102 $\gamma$-like events above the background (5144 $\pm$ 72) which corresponds to a significance of 0.8\,$\sigma$ calculated with formula 17 of \citet{LiMa1983} \footnote{The higher energy threshold of this analysis with respect to the previous ones (of this source) is caused by the fact that observations were performed at high zenith angle and part of them under moderate moonlight.}. The number of excess events and significances for the individual observation periods and for the complete 2011 data set are reported in Table~\ref{magicr}. Since none of the periods provided any significant detection, we compute two differential upper limits in the energy window from 125\,GeV to 500\,GeV, neglecting higher energies due to EBL absorption. 
The differential upper limits on the flux have been computed using the method of \citet{Rolke2005}, assuming a power law with a spectral index of 3.5 and a systematic error of 30\%. The results obtained are summarised in Table~\ref{ul} and in Figure~\ref{vhe_sed}, together with historical MAGIC observations, all corrected for EBL absorption using the model from \citet{Dominguez2011}. We have also computed the upper limits using 2.5 and 4.5 as spectral indices of the power law, and they do not differ appreciably from the values obtained using an index of 3.5.

\begin{table}
 \centering
\caption{ Differential upper limits calculated from MAGIC observations. Columns 1 and 2 give the energy and the respective absorption factor (a.f.) $e^{-\tau}$, where $\tau$ is the optical depth given by the EBL model of \citet{Dominguez2011}. In Cols. 3-5, the observed differential upper limits for the individual periods and the overall data sample are shown.}
\label{ul}
\begin{tabular}{ c c c  c c }
  \hline \hline
Energy &  a.f. & \multicolumn{3}{c}{ Upper Limit [$10^{-12}$ erg cm$^{-2}$s$^{-1}$] } \\
 $[$GeV$]$  &   & 2011 Feb- Apr & 2011 Jun & all 2011  data \\ 

\hline

  147.1 & 0.63 & 10.7 & 16.0 & 10.9 \\ 
  303.6 & 0.16 &  0.8 & 2.9  & 0.8 \\ 
  \hline
\end{tabular}
\end{table}

\section{Multiwavelength data}

  \begin{figure*}
  \centering
  \includegraphics[width=\textwidth,angle=270]{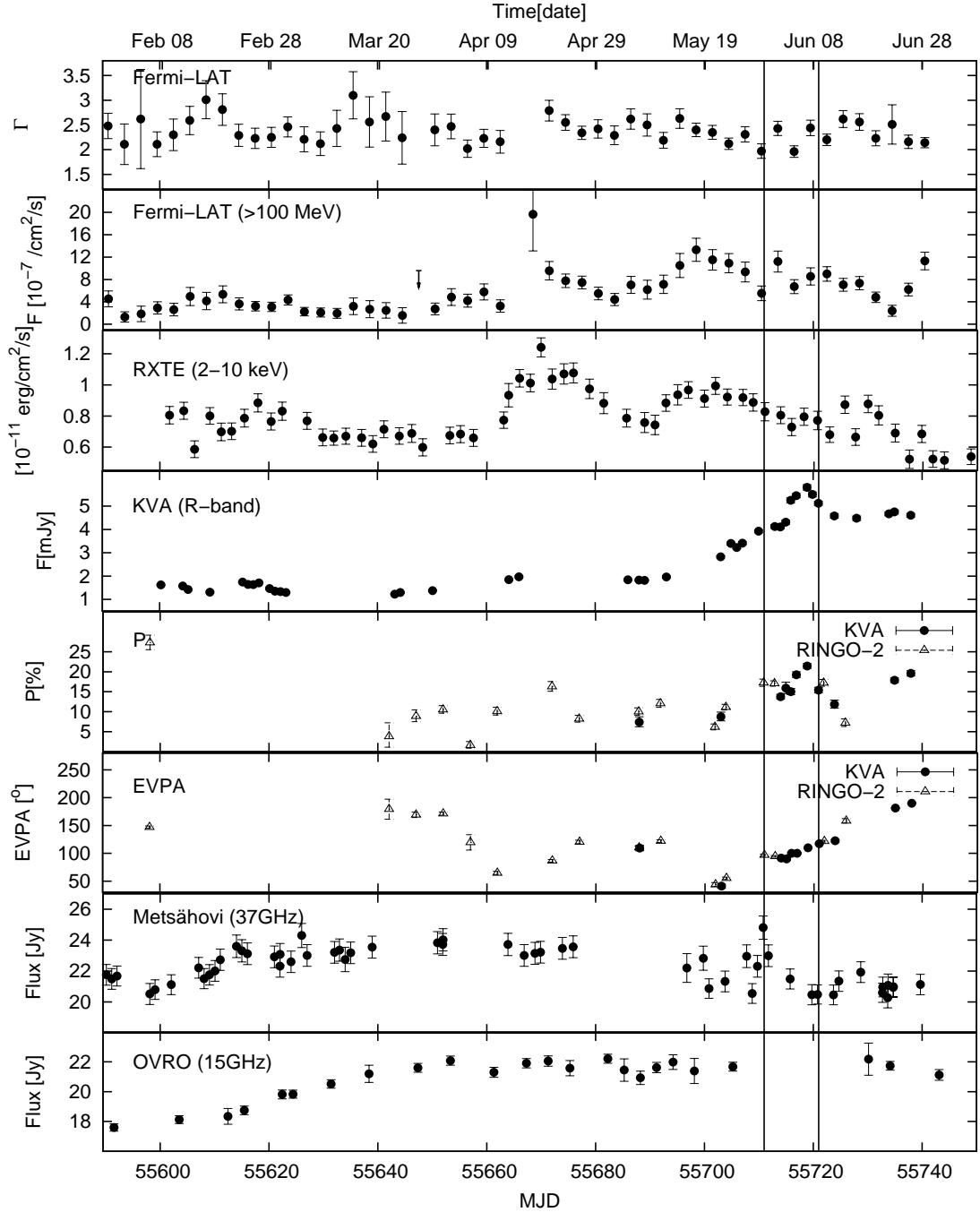}  
 \caption{Multiwavelength light curve from February 2011 to June 2011. The MAGIC ToO observation window is marked by vertical lines. Starting from the top panel: HE $\gamma$-ray observations from \lat (both flux and spectral index above 100\,MeV; the downward arrow for MJD~55646--55649 indicates a 95\% confidence level upper limit on the flux); X-ray data from RXTE; optical R-band photometric observations from the KVA telescope; optical polarisation measurements (both percentage of polarised flux and degree of polarisation) from the KVA (filled circles) and Liverpool telescopes (open triangles); and radio observations at 37\,GHz and 15\,GHz provided by the Mets\"ahovi and OVRO telescopes, respectively.}
\label{mw_lc}
  \end{figure*}
 
\subsection{HE $\gamma$-rays: \lat}

\lat is a pair-production telescope with a large effective area ($6500$\,cm$^2$ on axis for $>1$\,GeV photons) and a large
field of view (2.4\,sr at 1\,GeV), sensitive to $\gamma$-rays in the
energy range from $20$\,MeV to above $300$\,GeV~\citet{LAT}. 

Information regarding on-orbit calibration
procedures is given in \citet{Ackermann2012}. \lat normally operates in a
scanning ``sky-survey" mode, which provides a full-sky coverage every
two orbits (3 hours).
The analysis was performed following the \lat standard analysis
procedure\footnote{See details in
http://fermi.gsfc.nasa.gov/ssc/data/analysis/} using the \lat analysis
software, \textit{ScienceTools v9r29r2}, together with the
\textit{P7SOURCE\_V6} instrument response functions. 

The events were selected using SOURCE event class ~\citep{Ackermann2012}.  
We discarded events with zenith angles greater than $ 100^{\circ}$ and excluded time periods when the spacecraft
rocking angle relative to zenith exceeded 52$^{\circ}$ to avoid contamination by $\gamma$-rays produced in the Earth's atmosphere. 
The zenith angle is the angle between the event direction and the line from the centre of the Earth through the satellite. 

We selected events of energy between $100$\,MeV and $300$\,GeV 
within $15^{\circ}$ of the position of \object{3C~279}. Fluxes and
spectra were determined by performing an unbinned maximum likelihood fit
of model parameters with {\tt gtlike}. We examined the significance of the $\gamma$-ray signal from the sources by
means of the test statistic (TS) based on the likelihood ratio test\footnote{TS corresponds to $-2\Delta L = -2\log(L0/L1)$, where L0 and L1 are the maximum
likelihoods estimated for the null and alternative hypotheses, respectively.
Here for the source detection, $TS=25$ with 2 degrees of freedom corresponds
to an estimated $\sim 4.6\,\sigma$ pre-trial statistical significance assuming that the
null hypothesis TS distribution follows a $\chi^2$ distribution \citep[see][]{ML}.}.
The background model applied here includes standard models
for the isotropic and Galactic diffuse emission components\footnote{\texttt{iso\_p7v6source.txt} and \texttt{gal\_2yearp7v6\_v0.fits}}. In addition,
the model includes point sources representing all $\gamma$-ray emitters 
within the region of interest based on the Second \lat Catalog
\citep[2FGL:][]{2FGL}; flux normalisations for the diffuse and point-like 
background sources were left free in the fitting procedure.
Photon indices of the point-like background sources within
5$^{\circ}$ of the targets were also set as free parameters.
Otherwise the values reported in the 2FGL Catalog were used.

We derived a light curve in the \lat HE band using three-day time bins (Figure~\ref{mw_lc}). We plotted 95\% confidence level upper limits 
where the time bin has a TS $<$10. We note that the exposure times for \object{3C~279} in observations
between MJD~55646 and 55649, and between MJD~55664 and 55671
were significantly reduced (5-10 times shorter than usual)
because of ToO pointing-mode observations of Cyg\,X-3 and Crab Nebula, respectively.
In particular, the upper limit at MJD~55664--55667 corresponds to $6.2\times10^{-6}\,$ cm$^{-2}$ s$^{-1}$,
which is far beyond the range of the LAT light curve panel. 
The source was in a relatively low state at the beginning of the year,
followed by a period of enhanced activity. The light curve shows two
flares, with the peaks around MJD~55670 and 55695, and reaches a maximum HE flux of about $13 \times
10^{-7}\,$ cm$^{-2}$ s$^{-1}$, corresponding to roughly half the flux
level of the outburst measured in 2009 February \citep{2012Hayashida}.
Although the result shows the highest flux level at MJD 55667--55670,
the point has a large error bar because of the short exposure time
for 3C\,279 during the ToO observation that coincided with the rising phase 
of the first flare. Interestingly (see next section), the X-ray light curve shows a
similar trend, with two subsequent flares, the first one being the more intense.

The $\gamma$-ray spectra of 3C\,279 were extracted using data for two
periods: (A) from 2011 February 8 to 2011 April 12 (MJD 55600 -- 55663)
and (B) from 2011 June 1 to 2011 June 8 (MJD 55713 -- 55720).  These
periods include the MAGIC observing windows. Each $\gamma$-ray spectrum
was modelled using simple power-law (${\rm d}N/{\rm d}E \propto
E^{-\Gamma}$) and log-parabola (${\rm d}N/{\rm d}E \propto
(E/E_0)^{-\alpha - \beta \log(E/E_0)}$) models, as done in the Second \lat
Catalog and in a previous study of the source \citep{2012Hayashida}. In the case of log-parabola model, the parameter $\beta$ represents the curvature around the peak.
Here, we fixed the reference energy $E_0$ at 300 MeV. The best-fit parameters
calculated by the fitting procedure are summarised in
Table~\ref{GammaSPfit}. For the spectrum in period A, a
log-parabola model is slightly favoured to describe the $\gamma$-ray spectral
shape over the simple power-law model with the difference of the logarithm of
the likelihood fits $-2\Delta L =6.0$, which corresponds to a
probability of $1.43\%$ for the power-law hypothesis, while there is no
significant deviation from the simple power-law model in the spectrum of
period B.

\begin{figure}
  \centering
\includegraphics[width=88mm]{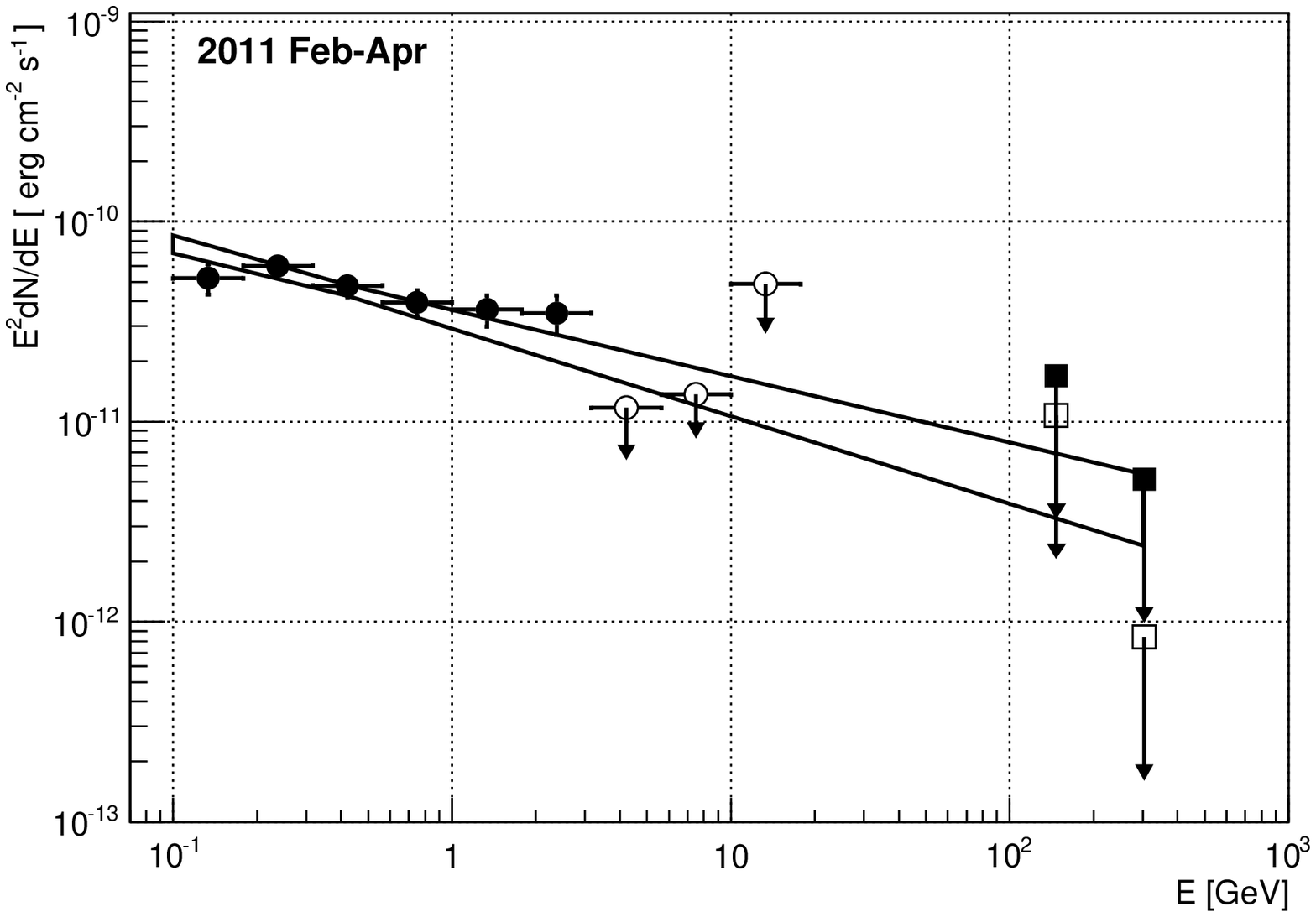}
\includegraphics[width=88mm]{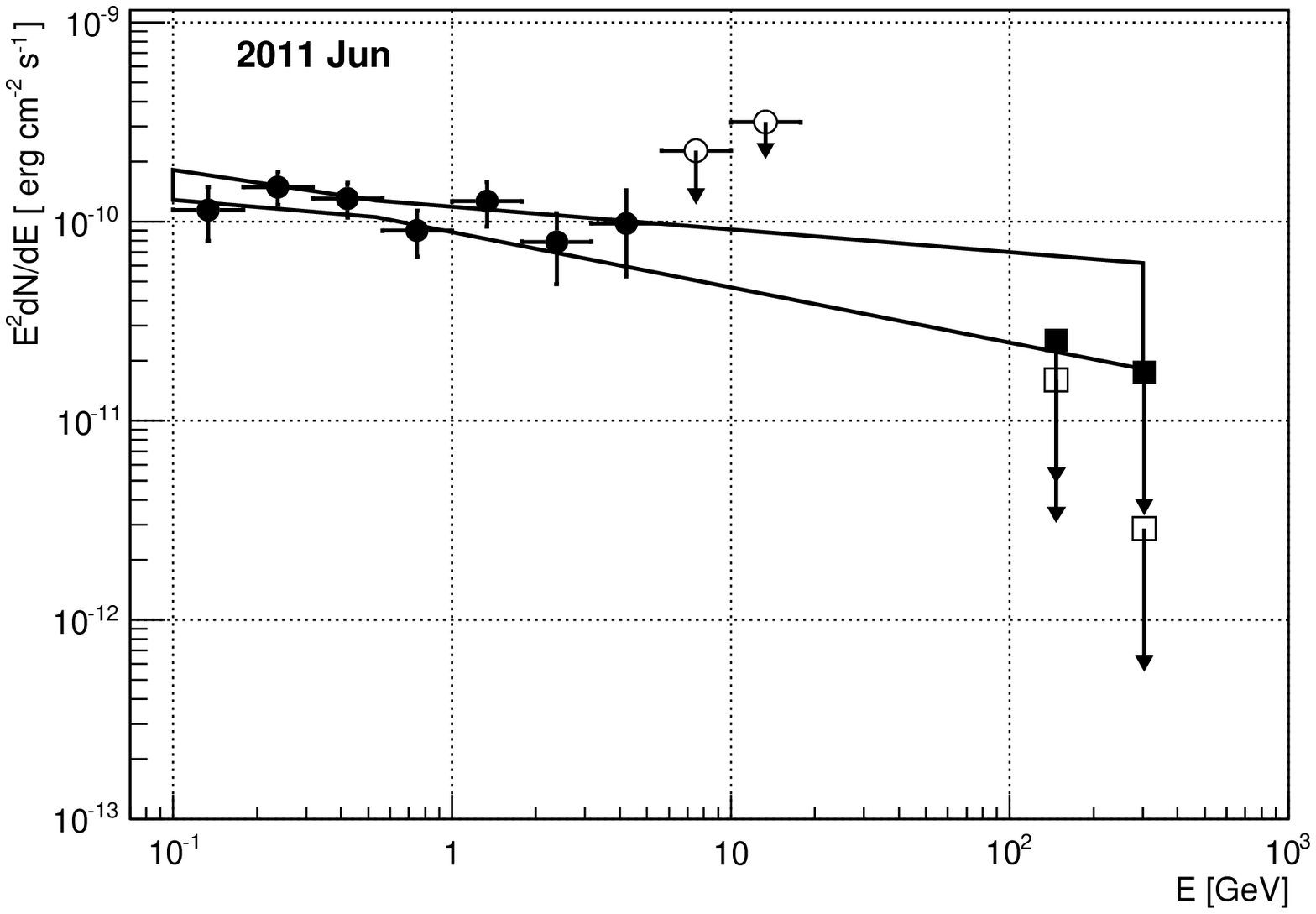}  
\caption{Comparison of \lat
observations and MAGIC upper limits for
2011 February-April (top) and 2011 June (bottom) observations. The butterfly
represents the SED plot with a $1 \sigma$ confidence region of the
best-fit power-law model extended up to 300\,GeV for \lat observations (filled circles, empty circles represent upper limits). The VHE data (open squares) are corrected for EBL absorption (filled squares) using \citet{Dominguez2011}.}
\label{fermi_magic}
  \end{figure}

In Figure~\ref{fermi_magic}, SED plots are shown together with a $1
\sigma$ confidence region of the best-fit power-law model for each
period, extended up to 300\,GeV. \lat data points and MAGIC upper
limits, both observed and corrected for EBL absorption, are also shown.
In both spectra, the detection significance of the \lat data ($TS \sim
400$) was not statistically sufficient for 3C\,279 to determine a
spectral break in the \lat data alone as previously obtained
\citep{2012Hayashida}. Considering period A, the VHE upper limits do not
indicate the presence of a break or a curvature between \lat and MAGIC
energy ranges. On the other hand, in the June spectrum, the MAGIC upper
limits points are located almost at the edge of the  $1 \sigma$
confidence region of the LAT spectral model, suggesting a break or a
curvature between the energy ranges of the two experiments. This
distribution is consistent with the spectrum reported in the Second \lat 
Catalog \citep{2FGL}, where a log-parabola model was used. Moreover, 
curvature was also reported in \citet{2012Hayashida}, where a larger data
sample (2 years) was used.

We also investigated the highest energy photons associated
with 3C\,279 during each period,
including some quality checks for each event: 
the tracker section in which the conversion occurred,
angular distance between the reconstructed arrival direction of the event and 3C\,279,
probability of association estimated using {\tt gtsrcprob}\footnote{The tool assigns the probabilities for each event including not only the spatial consistency, but also the spectral information of all the sources in the model. See details at  http://fermi.gsfc.nasa.gov/ssc/data/analysis/scitools/help/gtsrcprob.txt},
and whether the event survives a tighter selection than the standard {\tt source:evclass=2} selection.
The results are summarised in Table~\ref{lathephoton}.

\begin{table*}[!ht]
\centering
\caption{Results of spectral fitting in the HE $\gamma$-ray band
measured by \textit{Fermi}--LAT. Column 1 shows the period of observations,
Cols. 2-4 the fitting model and its parameters (PL: power-law
model, LogP: log-parabola model. See definitions in the text), Col. 5 the TS, Col. 6 the difference of the logarithm of the
likelihood of the fit with respect to a single power-law fit. The flux
$>100$\,MeV is given in the last column. \label{GammaSPfit}}
\begin{tabular}{ccccccc}
\hline  \hline
Period & \multicolumn{5}{c}{Gamma-ray spectrum (\textit{Fermi}--LAT) } & Flux ($>100
$\,MeV) \\
(MJD) & fitting model & $\Gamma/\alpha$ & $\beta$ &   $TS$ & $-2
\Delta{L}$ & $ (10^{-7}$ ph cm$^{-2}$ s$^{-1}$)  \\
\hline
2011 Feb 8 -- Apr 12   &   PL   &    $2.37\pm0.06$ &  &    695.6 &  & $3.5
\pm0.3$ \\
(55600 -- 55663)  & LogP  & $2.18\pm0.10$ & $0.12\pm0.06$ &   700.9  & 5.3 & $3.2\pm0.3$ \\
2011 Jun 1 -- Jun 8   &    PL  &  $2.17\pm0.08$ & &   400.6 &  & $8.3\pm1.0$
\\
(55713 -- 55720)  & LogP  & $2.02\pm0.15$ & $0.07\pm0.06$ &   402.0 &  1.4 & $7.7\pm1.0$ \\
\hline
\end{tabular}\\
\end{table*}

\begin{table*}
 \centering
 \caption{Highest energy photons associated with \object{3C~279} in \lat
observations during each period of the MAGIC observations in 2011.}
\label{lathephoton}
 \begin{tabular}{c c c c c c c c }
   \hline\hline
  &    & estimated & incident & converted & reconstructed  &  probability of  & survived   \\
 &  detection time of the event  & energy & angle & layer & arrival direction  & association  &  tighter event
\\
&    & [GeV] & &  & from 3C\,279\tablefootmark{a}    & with
\object{3C~279} &   selection\tablefootmark{b}\\
 \hline
Period A  & May 10 (MJD 55691.2821) & 19.8 & $51.4^{\circ}$ & back &
$0.11^{\circ}$   & 98.9\% & yes \\ \hline
Period B &  June 05 (MJD 55717.0275) &  13.1 & $45.2^{\circ}$ &back &
$0.17^{\circ}$ &  98.4\% & yes \\
\hline
 \end{tabular}

\tablefoottext{a}{68\% containment radius
of the LAT point-spread function is $0.304^{\circ}$
in the instrument response functions of \textit{P7SOURCE\_V6} for the back-thick layers converted events at 33.5
\,GeV with an incident angle of $47.0^{\circ}$.} \\
\tablefoottext{b}{so-called ultraclean:evclass=4 data selection.} \\

 \end{table*}

\subsection{X-rays: RXTE-PCA}

The source \object{3C~279} has been monitored with the Rossi X-Ray Timing Explorer (RXTE) since 1996 \citep{Chatterjee2008}. It has been observed with the PCA instrument in separate pointings with a typical interval of two to three days and exposure times of the order of kiloseconds. For the analysis, routines from the X-ray data analysis software FTOOLS and XSPEC were used. The source spectrum from 2.4 to 10 keV is modelled with a power law with a low-energy photoelectric absorption by the intervening gas in our Galaxy, which is represented by a hydrogen column density of  8$\,\times\,10^{20}\,$atoms\,cm$^{-2}$ (Chatterjee et al. 2008).

Compared to the long-term X-ray behaviour of \object{3C~279} in 1996-2007 presented in \citet{Chatterjee2008}, the source was in a low state during spring 2011 (Figure~\ref{mw_lc}). The light curve in the energy range 2-10\,keV shows two minor flares peaking around MJD 55670 (with a maximum flux of $1.2\times10^{-11}\,$erg\,cm$^{-2}$\,s$^{-1}$) and MJD 55700 (with a maximum flux of $1.0\times10^{-11}\,$erg\,cm$^{-2}$\,s$^{-1}$). For comparison, the major X-ray flares of the source have reached peak fluxes of $(3-5)\times10^{-11}\,$erg\,cm$^{-2}$\,s$^{-1}$, while the \lat outburst reported in \citet{Abdo2010Nature} had a maximum flux of similar order to the outbursts reported here. The bowties represented in the multifrequency SEDs  (see Sect. \ref{subsect:SED_FA}) are obtained using the flux $(6.7\pm0.5)\times 10^{-12}\,$erg cm$^{-2}$s$^{-1}$ and the energy index $0.6\pm0.2$ for the February to April observations and flux $(7.9\pm0.5)\times 10^{-12}\,$erg cm$^{-2}$s$^{-1}$ and the energy index $0.9\pm0.2$ for the June observations. 

\subsection{Optical observations: KVA and Liverpool}

The optical observations were performed with the Kungliga Vetenskapsakademien (KVA) telescope and the Liverpool telescope (LT), both located on the Canary Island of La Palma. 
The KVA telescope, operated remotely from Finland, consists of two telescopes mounted on the same fork; a 35\,cm Celestron and a 60\,cm Schmidt reflector. Photometric monitoring of \object{3C~279} has been performed regularly since 2004 as a part of the Tuorla blazar monitoring program\footnote{http://users.utu.fi/kani/1m/}. Observations were performed with the KVA 35\,cm telescope in the R-band and data were analysed using standard procedures \citep[for details see][]{Aleksic2011a}. The R-band light curve shows a constant quiescent state \citep[flux and magnitude of the source quiescent state are 1.45\,mJy and R=15.8; respectively][]{Reinthal2012} from 2011 February until May, around MJD 55700, after which there is a clear outburst. It reaches its peak at MJD 55719, showing a maximum flux of $\sim6$\,mJy (magnitude $R\sim14.3$). Compared to previous outbursts observed from this source, it is the third brightest since the beginning of the program in 2004. The other two brighter optical outbursts were detected in  February 2006 and January 2007 in coincidence with the detections at VHE $\gamma$-rays \citep{Albert2008Science, Aleksic2011a}.

The polarisation monitoring of \object{3C~279} had been carried out since 2009 using the KVA 60\,cm telescope, equipped with a CCD polarimeter capable of polarimetric measurements in BVRI bands using a plane-parallel calcite plate and a super-achromatic $\lambda/2$ retarder \citep{Piirola2006}. The observations presented here were performed without a filter. Since 2010, polarimetric observations have also been conducted with the fully-robotic 2\,m LT. For the present campaign\footnote{This campaign is part of a larger program conducted at the LT that provides optical polarisation data to complement MAGIC VHE observations of extragalactic sources.}, polarimetry data-taking was intensified in June 2011, 
triggered by the high-activity state detected in the \lat and optical bands. The source was followed for about a month, with an almost daily observation frequency. At the epochs of intense 
monitoring during the high-activity state in MJD 55710-55730, we were able to closely follow the smooth evolution of the polarisation parameters which allowed us to model in detail the behaviour of the source (see Sect.4.2).

Observations at the LT were performed with the novel RINGO2 fast-readout imaging polarimeter \citep{Steele2010}, equipped with a hybrid V+R filter, 
consisting of a 3\,mm Schott GG475 filter cemented to a 2\,mm KG3 filter. The polarimeter used a rotating polaroid with a frequency of 
approximately 1\,Hz, during the cycle of which eight exposures of the source are obtained. These exposures were synchronised with the phase of the polaroid to determine the degree and angle of polarisation. The flux of \object{3C~279} was measured in each of the eight images using aperture
photometry, and the normalised Stokes parameters $q=Q/I$ and $u=U/I$ \citep[for a definition of the Stokes parameters
see e.g.][]{Rybicki1986} and 
their errors $\sigma_q$ and $\sigma_u$ were computed using the formulae 
in \citet{Clarke2002}. The RINGO2 instrument exhibits 
an instrumental polarisation which remains constant through several 
months, but changes abruptly at epochs when changes have been made to 
the system. This instrumental polarisation amounts to 2.1-2.5\% 
depending on which interval is considered. To correct for instrumental 
polarisation, the average $q$ and $u$ of zero polarisation standards, 
monitored throughout the \object{3C~279} campaign, were subtracted from the $q$ 
and $u$ of \object{3C~279} and the errors propagated into $\sigma_q$ and 
$\sigma_u$. The degree of polarisation $p$ and the EVPA were then computed from $p = \sqrt{q^2 +u^2}$ and EVPA = $0.5 
\tan^{-1}(u/q)$. After this, the unbiased degree of polarisation $p_0$ 
was computed from $p_0 = \sqrt{p^2-\sigma^2}$ where $\sigma$ is 
$(\sigma_q + \sigma_u)/2$. The (asymmetric) 68\% and 95\% confidence 
intervals of $p_0$ were then computed using the prescription by \citet{Simmons1985} and, if the lower 95\% confidence boundary 
of $p_0$ was $> 0$\%, we considered that significant polarisation had 
been detected. In this case a correction for instrumental 
depolarisation, determined from high polarisation standards, was applied 
using $p_{corr} = p_0 / k$, where $p_{corr}$ is the corrected 
polarisation and $k = 0.76 \pm0.01$. In the corresponding panels of Figure~\ref{mw_lc}, $p_{corr}$ and its 68\% confidence limits are given. Finally, the error of the 
EVPA was computed from $\sigma_{EVPA} =28.65\, \sigma_{p_{corr}} / p_{corr}$.

\subsubsection{Polarisation}
 
\begin{figure}
  \centering
  \includegraphics[height=88mm, angle=270,clip]{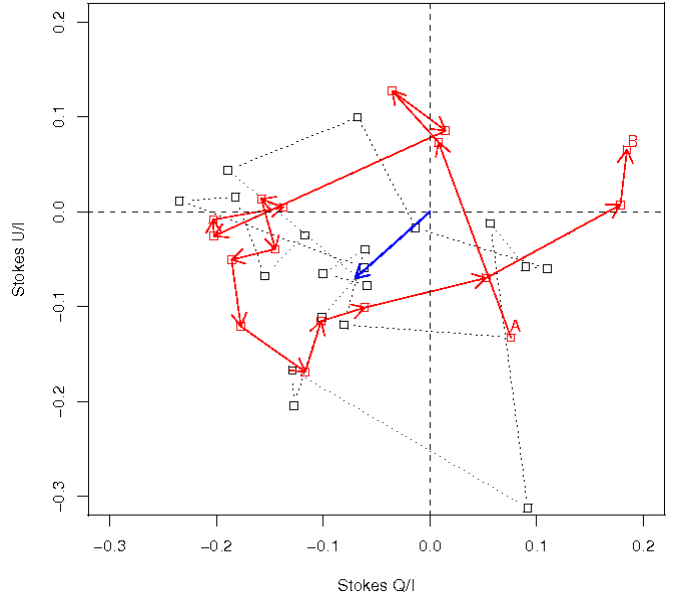}
 \caption{Stokes plane of the polarisation data, where each coordinate axis represents one of the orthogonal components of the linearly polarised light. The distance to the coordinate centre is the polarisation degree. Since the polarisation is a pseudo-
vector, with a $\pi$-ambiguity, the reported values for the EVPA rotations correspond to half-angles in the Stokes plane. Open boxes show the KVA and RINGO2 polarisation data for \object{3C~279}, from February to June 2011: the grey boxes are from the period prior to optical flare, while the red open boxes are from the time of optical flare (starting $\sim$MJD 55700). The lines (grey dotted for pre-flare epoch and red arrows for flare epoch)
connect the points in chronological order. The blue vector marks the average direction and magnitude of the polarisation of the
source measured during the campaign. }\label{stokes} 
  \end{figure}

The complete light curves for the polarisation degree and EVPA are presented in the corresponding panels of Figure~\ref{mw_lc}. The plots show that the optical flare was accompanied by a fourfold increase in the degree of polarisation. The polarisation 
degree reached a maximum of $\sim20\%$ contemporaneously with the photometric peak flux, after which it returned to the initial value of $\sim5\%$, while the total flux remained persistently high for the remainder of the monitoring. Throughout the observation period the polarisation was variable, showing high-amplitude changes at timescales of a few days. High polarisation such as that reached at the maximum of the flare is typical during high-activity states of the source and reflect a high degree of local ordering of the magnetic field \citep[e.g.][]{Larionov2008}. The temporal coincidence of the photometric and polarised flares suggests that the two events are related.

The EVPA also presents notable evolution. During the approximately 30 days spanning the optical high-activity state, observed with high cadence, the polarisation position angle smoothly rotated $\sim140^\circ$ with a nearly constant rate. The rotation showed an inversion of the sense of rotation in the mid-point of the event, when the polarisation degree decreased to a local minimum. Since the monitoring was discontinued soon after the 
optical flux started decreasing from its maximum, it is not possible to judge if the rotation reached its end point during the observations.
Outside of the epochs of smooth rotation, the direction of the EVPA varied erratically, suggesting that there was no single polarised component dominating the source evolution at all observed epochs. 

Figure~\ref{stokes} shows the Stokes plot for every polarisation measurement presented here and provides an alternative visualisation of the polarisation state of the source. The red points and arrows mark the chronological evolution of the rotation event (from A $\rightarrow$ B) observed during the optical flare, while the grey boxes are from the period prior to optical flare. The blue vector in the third quadrant is the mean polarisation vector, averaged over all values measured during the campaign. We note that the mean EVPA, whose value is $\sim 20^\circ$, is nearly perpendicular to the jet position angle\footnote{An angular variation of $\Delta\theta$ in the Stokes Q-U plane corresponds to a change in the EVPA of $2\Delta\chi$; therefore, the  $360^\circ$ of the Stokes plane correspond to only $180^\circ$ of real angle $\chi$ because of the  $180^\circ$ ambiguity of the polarisation angle.}, in agreement with what is observed from the core position in the 43\,GHz VLBA images, suggesting that the dominant polarised optical emission was co-spatial to the radio core (Figure~\ref{vlba}) $\theta \sim 2^{\circ}$. The EVPA of the source randomly oscillated around this direction with a variance of $\Delta\chi \approx 20 ^{\circ}$. 

The polarisation variability outside the optical high-activity state can be described as magnetic turbulence around the mean field position. To evaluate whether the rotation event can also be described by magnetic turbulence, we have performed Monte Carlo simulations using a random ensemble of polarisation angles as a proxy for turbulence. We modelled the emitting region as $N$ cells of equal volume and field intensity. The field is uniform within the cell, but randomly oriented and acquires at each simulation step a new random value. The number of cells was determined from the mean fractional polarisation, assuming that each cell individually emits with the maximum degree of polarisation for incoherent synchrotron radiation, $P_{\rm{max}}\approx 0.7$. With an observed variance $\sigma_{\rm{P}}^2 \simeq 0.003$, we have $N = \sqrt{<P>/\sigma_{\rm{P}}} \sim 10$ cells. The net polarisation was then obtained from the superposition of the emission of the individual cells. The probability that plasma turbulence generates a continuous rotation of the EVPA by  $140^\circ$ after 15 epochs is $\lesssim 1$\%. It should be noted that the probability increases if we approximate the turbulence as N cells of equal volume, where about ten cells leave and around ten cells enter the emission region \citep{DArcangelo2008}.
While the random probability is not small enough that it could be safely excluded as a candidate mechanism to explain the EVPA rotation,  the smoothness of the rotation showing only small excursions from the prescribed path during its entire duration (compared to the $20^{\circ}$ variance of the rest of the campaign), does not favour turbulence as the likely explanation\footnote{After the submission of this paper \citet{Kiehlmann2013} presented larger optical polarization data set from several instruments, which supports this conclusion.}. Therefore we investigate a geometric effect as the possible cause of the rotation (see Sect.4.2.).

\subsection{Radio: VLBA 43\,GHz, Mets\"ahovi 37\,GHz, and OVRO 15\,GHz observations}

\begin{figure}
  \centering
 \includegraphics[width=88mm,trim=0cm 0cm 3cm 0cm, clip]{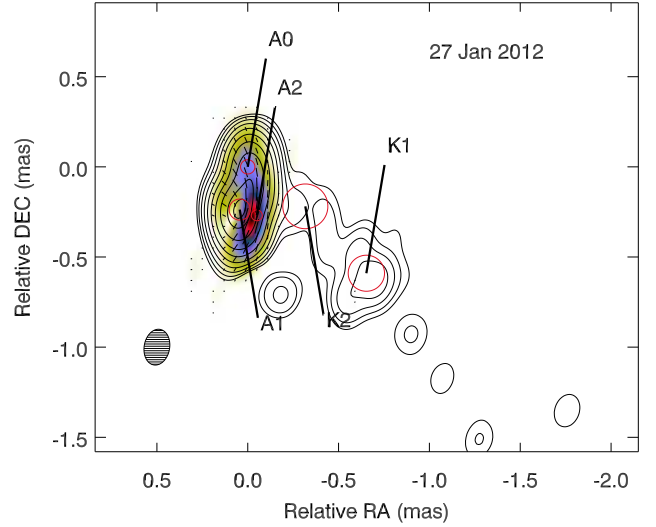}
 \caption{VLBA total (contours) and polarised (colour scale) images of \object{3C~279} at 43\,GHz with the total intensity peak of 17.0\,Jy/beam, polarised intensity peak  of 0.80\,Jy/beam, and a Gaussian restoring beam$=0.13\times0.20\,$mas$^2$ at $PA=-6^\circ$; contours represent $0.25,0.5,\dots, 64\%$ of the peak intensity; line segments within the image show direction of linear polarisation; red circles indicate position and size (FWHM) of components according to model fits. Four bright components are moving within 1\,mas from the core (A0), but there is no ejection of new components related to the flaring episodes reported in this work.} 
 \label{vlba}
  \end{figure}

\begin{figure}
  \centering
 \includegraphics[angle=270,width=88mm,clip]{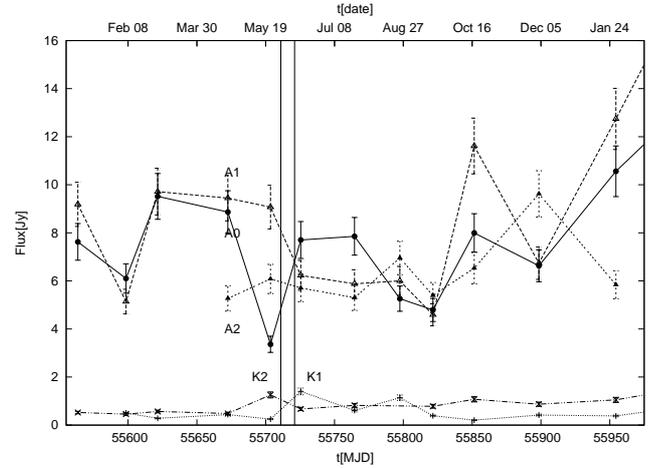}
   \caption{ The fluxes of the single components of the VLBA 43 GHz core between January 2011 and January 2012. The MAGIC ToO observation window is marked by vertical lines. There was no new component ejected from the VLBA core in 2011; the brightest components are the core (A0) and the innermost component (A1). } 
\label{vlba2}
  \end{figure}

\begin{figure}
  \centering
 \includegraphics[angle=270,width=88mm,clip]{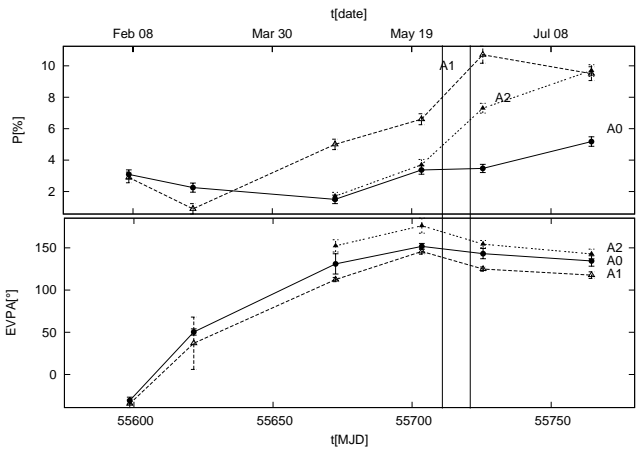}
   \caption{ VLBA polarisation data of the single components between February 2011 and July 2011. The MAGIC ToO observation window is marked by vertical lines. The EVPA, of both components A0 and A1, rotates with $\sim180^{\circ}$ between January and March 2011. However, by the time the rotation starts in the optical regime the EVPA of the VLBA core (and the components closest to it) has stabilised to $\sim150^{\circ}$.} 
\label{vlba_pol}
  \end{figure}

The source \object{3C~279} was monitored by the Mets\"ahovi radio observatory, the Owens Valley Radio Observatory (OVRO), and the Very Long Baseline Array (VLBA) as a part of quasar monitoring programs.

The 37\,GHz observations were performed with the 13.7\,m diameter Mets\"ahovi radio telescope, a radome-enclosed paraboloid antenna situated in Finland. The measurements were made with 
a 1\,GHz-band dual beam receiver centred at 36.8\,GHz. The high electron mobility pseudomorphic transistor front end operates at room temperature. The observations are ON-ON observations, alternating the source and the sky in each feed horn. The flux density scale is set by observations of DR 21 (a huge molecular cloud located in the constellation of 
Cygnus, the standard candle for radio astronomy). The sources NGC~7027, 3C~274, and 3C~84 are used as secondary calibrators. A detailed description of the data reduction and analysis can be found in \citet{Teraesranta1998}. The error estimate in the flux density includes the contribution from 
the measurement RMS and the uncertainty of the absolute calibration.

Regular 15\,GHz observations of \object{3C~279} were carried out using the OVRO 40\,m telescope \citep{Richards2011}. This program commenced in late 2007 and now includes about 1800 sources, each observed with a nominal biweekly cadence. The OVRO 40\,m uses off-axis dual-beam optics and a cryogenic high electron mobility transistor low-noise amplifier with a 15.0\,GHz centre frequency and 3 GHz\,bandwidth. The total system noise temperature, including 
receiver, atmosphere, ground, and CMB contributions, is about 52\,K. The two sky beams are Dicke switched using the off-source beam as a reference, while the source is alternated between the two 
beams in an ON-ON mode to remove atmospheric and ground contamination. A noise level of approximately 3--4\,mJy in quadrature with about 2\% additional uncertainty, mostly due to pointing 
errors, is achieved in a 70\,s integration period. The calibration is performed using a temperature-stable diode noise source to remove receiver gain drifts; the flux density scale is derived 
from observations of 3C~286 assuming the  value of 3.44\,Jy at 15.0\,GHz \citep{Baars1977}. The systematic uncertainty of about 5\% in the flux density scale is not included in the error 
bars. Complete details of the reduction and calibration procedure can be found in \citet{Richards2011}.

Both the 37\,GHz and 15\,GHz radio light curves (Figure~\ref{mw_lc}) show a smooth increase at the beginning of the observation period after which the flux stays constant. The 15\,GHz light curve 
shows a gap in coincidence with the optical outburst, while the flux of the 37\,GHz light curve increases from 20\,Jy to 24\,Jy during the rising phase of the optical outburst. The flux increase resembles a flare, with a sharp rise and a slower decay; the peak flux is reached at MJD 55710, i.e. 10 days before the peak of the optical outburst.

The VLBA observations at 43\,GHz have been performed once a month since the beginning of the monitoring program, in 2007\footnote{http://www.bu.edu/blazars/VLBAproject.html}. 
The data were analysed as described in \citet{Jorstad2005}, extracting the information about the jet kinematics and polarisation in the period from January 2011 to January 2012. 
In this period there were four bright moving components within 1\,mas of the core (see Figure~\ref{vlba}), but no new components ejected from the core (the latest injections are A1 and A2 with zero separation epochs MJD  55281.4 and 55335.84). The fluxes of the single components are shown in Figure~\ref{vlba2}. From February 2011, the core (A0) and the component very close to the core (A1) were the 
brightest components. During the optical outburst (starting at MJD 55700) the core flux increased, while for A1, A2 and K1 the flux decreased. The polarisation data for the period between February 2011 and July 2011 show that the EVPA of the VLBA core at 43\,GHz is constant at $150^\circ$ between  MJD $\sim$ 55700 and $\sim$55750, a period which includes the optical rotation, but because of the time resolution a rotation of $180^\circ$ cannot be excluded between these two epochs.

The polarisation data of the components A0, A1, and A2 are shown in Figure~\ref{vlba_pol}. Between January and March 2011, the EVPA of the components A0 and A1 rotates with $\sim180^\circ$, but by the time the rotation starts in the optical, the EVPA of the VLBA core (and the components closest to it) has stabilised to  $\sim150^\circ$. 
This behaviour resembles the profile and the characteristics observed in the core at 43\,GHz reported in \citet{Larionov2008}, with the only 
difference being the absence of a simultaneous rotation in the optical regime. 

\section{Discussion}

We now discuss plausible emission scenarios for the two MAGIC observation periods and examine in detail a geometric interpretation of the observed rotation of the optical EVPA.

\begin{figure}
  \centering
\includegraphics[width=88mm]{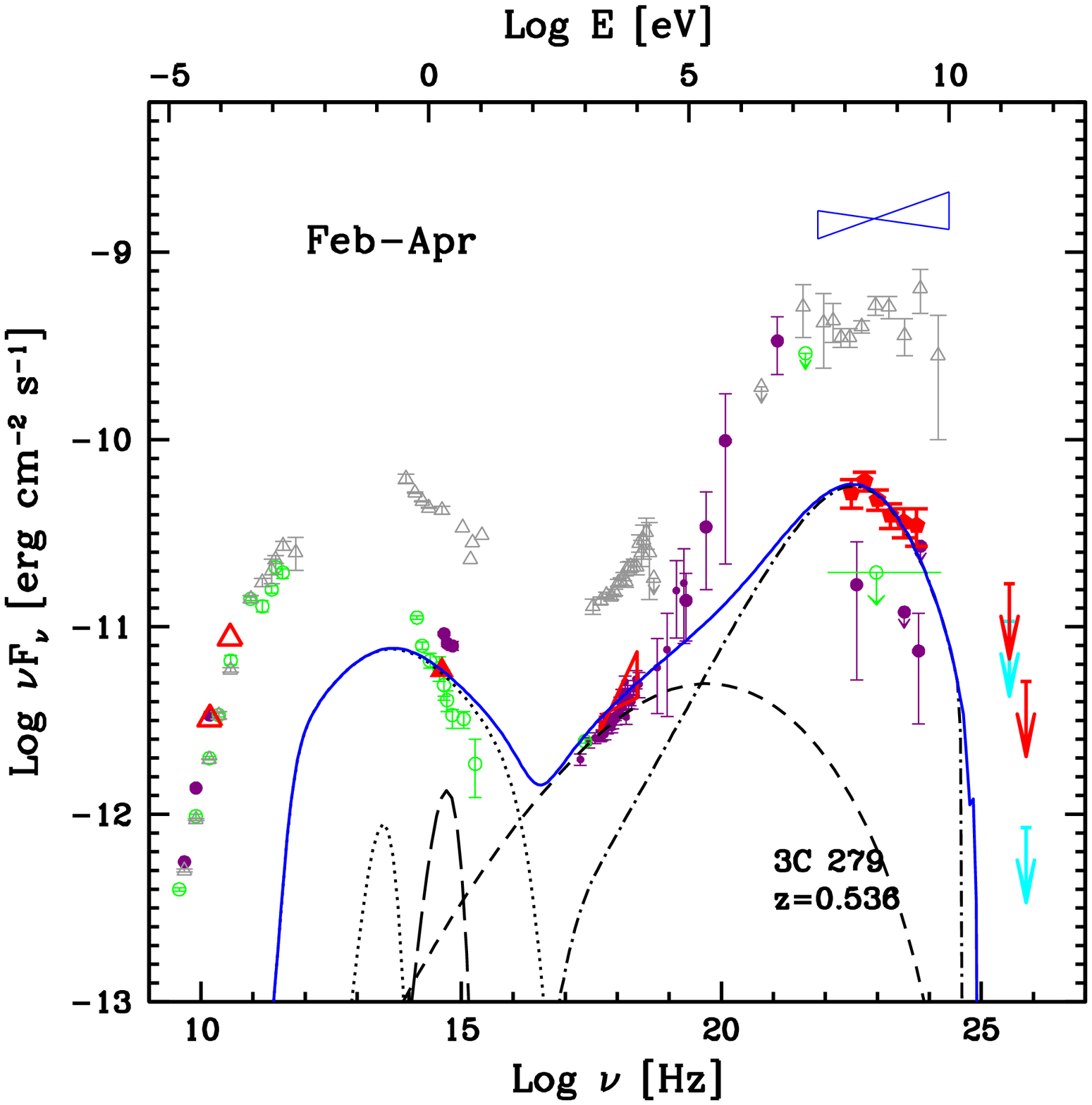}
  \includegraphics[width=88mm]{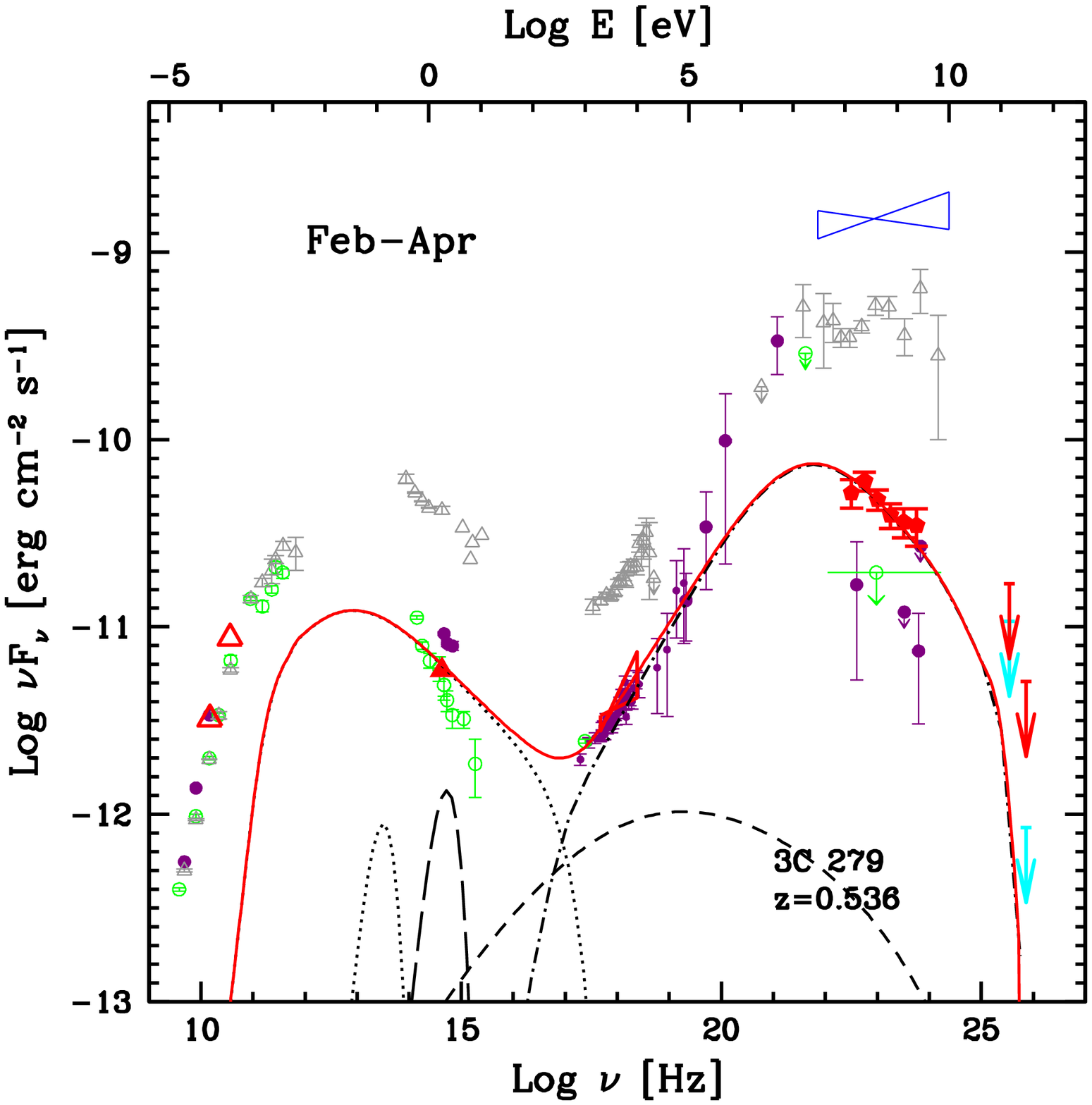} 
\caption{Multiwavelength observations between February and April 2011:
radio from Mets\"ahovi and OVRO (red open triangles), optical from KVA (red filled triangles), X-ray from RXTE (red bowtie), and HE $\gamma$-rays from \lat (red circles) and MAGIC (red arrows: EBL corrected, cyan arrows: observed points).
Historical data are also shown \citep{Aleksic2011a}: high-activity state from 1991 \citep[grey open triangles,][]{Hartman1996}, low state from 1993 \citep[green open circles,][]{Maraschi1994,Ballo2002}, the high-activity state from 1996 \citep[blue bowtie,][]{Wehrle1998}, and the low-activity state from the end of 1996 to the beginning of 1997 \citep[violet dots][]{Hartman2001}. Only points marked in red are considered for the SED fit. The SED is fitted using two leptonic models, with different populations of external low-energy photons as target for the inverse Compton process. The individual components are shown: synchrotron (dotted), SSC (dashed), and EC (dot-dashed). The black-body radiation from the BLR (dashed) and from the infrared torus (dotted) are also shown. In the upper panel, the high-energy emission originates from a region located inside the BLR. In the lower panel, the high-energy emission comes from a region located outside the BLR, considering only photons from the infrared torus as targets for inverse Compton scattering. The parameters are summarised in Table~\ref{fitsed}.}
\label{sed_fa} 
  \end{figure}

\subsection{Multifrequency variability and spectral energy distributions}

We investigate the multifrequency behaviour during the two MAGIC observing periods to constrain the number and location of the emission regions and then model the spectral energy distributions of these epochs accordingly. We have compiled the multiwavelength spectral energy distribution for the 
two periods of observations, February-April 2011 (Figure~\ref{sed_fa}) and June 2011 (Figure~\ref{two_zones}). 

\subsubsection{February-April 2011: low state}
\label{subsect:SED_FA}

In the first period the source was in a rather low state in optical to $\gamma$-ray wavebands (see Figure~\ref{mw_lc}), while some activity was reported in the radio bands. The light curves at 15\,GHz and 37\,GHz showed an increasing flux, with the 37\,GHz light curve peaking before the 15\,GHz light curve. At 43\,GHz, the flux density of the central components increases, while no significant variability is observed in the components K1 and K2 (Figure~\ref{vlba2}). This may suggest that the variability is related to the central region ($< 1$\,mas), but the lack of variability in the other bands does not allow us to draw strong conclusions on the site of the emission.

The multifrequency SED is fitted using leptonic models \citep[for details see][]{Maraschi2003}. 
The emission region, a spherical blob with radius $R$ is filled with  a homogeneous and tangled magnetic field $B$ and a population of relativistic electrons. The spectrum of the electrons extends from $\gamma_{min}$ to $\gamma_{max}$ and is described by $N(\gamma)=K \gamma^{-n_1}(1+\gamma/\gamma_b)^{n_1-n_2}$. The electrons emit synchrotron radiation forming the first peak (dotted lines). This low-energy radiation is then inverse Compton upscattered to high energies (dashed lines), contributing to the second peak. In addition to the low-energy photons coming from the synchrotron process, two other populations of low-energy photons are considered as targets for the inverse Compton process (dot-dashed lines). The first one (blue curve in the upper panel of Figure~\ref{sed_fa}) is photons from the BLR, characterised by  $L_{disc}= 3 \times 10^{45}\,$ erg s$^{-1}$,  $R_{BLR}= 1.7 \times 10^{17}\,$ cm. The model also accounts for the internal pair absorption, considering the external BLR photons not only for the inverse Compton scattering, but also as target for $\gamma \gamma$ pair production.  In the other scenario (red curve in the lower panel of Figure~\ref{sed_fa}), photons stem from the infrared torus, whose luminosity and distance are, respectively, $L_{torus} = 2 \times 10^{45}\,$ erg s$^{-1}$ and  $R_{torus}= 2 \times 10^{18}\,$ cm. We assume an accretion disc as described in \citet{Pian1999} and that the infrared torus intercepts a fraction ($\sim60\%$) of the disc luminosity and re-radiates it in the form of a black-body spectrum with a temperature of $900\,K$ \citep{Calderone2012}. 

Figure~\ref{sed_fa} depicts the resulting SED fits whose parameters are listed in Table~\ref{fitsed}. Both scenarios can fit reasonably well the multifrequency data, and the obtained parameters have values within the typical ranges used for this source. Consequently, we cannot constrain the location of the emission region in this period by means of the SED modelling.   

The radio observations are not included in our SED fits, but are considered only as upper limits. The observed radio emission is assumed to originate from a different emission region, with main contributions from the VLBA core and parsec scale jet. These emission regions are at greater distances from the central engine \citep[e.g.][]{Pushkarev2012} and suggest that the VLBA core is located at $>7.88$\,pc from the central engine based on core shift measurements.

The modelled emission regions are assumed to be much closer to the central engine. To check this assumption, we can derive the maximum size of the emission region that will be synchrotron self-absorbed at radio frequencies. Following the approach by \citet{Abdo2010Nature} and adopting the flux at 43\,GHz to be $\sim 6\,$Jy, we calculate that  the emitting region must be smaller, in transverse size, than $R= 3.68 \cdot10^{17}$\,cm $=0.12\,$pc for a magnetic field $B = 0.8$\,G and a Lorentz factor of 10 in order to be optically thick at 43\,GHz and lower frequencies. This is in agreement with the radius derived in the SED fit (see Table~\ref{fitsed}).

The distance of the emission region from the central engine we derived is either larger than or comparable to previous works. \citet{Abdo2010Nature}, because of the absence in variations of the radio data up to 230\,GHz, position the emission region upstream from the radio core; \citet{Larionov2008} locate the emission region within $\sim 1\,$pc of the radio core. Given that the VLBA angular size of the core at 22\,GHz is $0.1\,$mas \citep{Wehrle2001}, the projected linear size for the location of \object{3C~279} is $\sim 0.6\,$pc ($1.8\times 10^{18}\,$cm). It follows that the active region is a fraction ($80\%$) of the core size or the transversal size of the pc-scale jet and the location of the active region is upstream from the core.

 \begin{table*}
 \centering
 \caption{Model parameters used for fitting the SEDs with different leptonic scenarios (Figs.~\ref{sed_fa} and ~\ref{two_zones}). The accretion disc and the torus are characterised by $L_{disc}= 3 \times 10^{45}\,$ erg s$^{-1}$, and $R_{BLR}= 1.7 \times 10^{17}\,$ cm, $L_{torus} 2 \times 10^{45}\,$ 
erg s$^{-1}$,  and $R_{torus}= 2 \times 10^{18}\,$ cm, respectively. $\delta$ is the Doppler factor and $\Gamma$ is the Lorentz factor; see the text for a description of the other parameters.}
\label{fitsed}
 \begin{tabular}{c c c c c c c c c c c c }
   \hline\hline
  &   & $\gamma_{min}$ &  $\gamma_{b}$ & $\gamma_{max}$ &  $n_1$ & $n_2$ & B [G] & K [$\mathrm{cm}^{-3}$] & R [cm] & $\delta$  & $\Gamma $ 
\\ \hline
\multirow{2}{*}{2011 Feb - Apr} &inside the BLR &  1    &   610    &    1.1$\times 10^4$    &  2   &    3.7   &     2.4   &      5.9$\times 10^5$      
&      4.7$\times 10^{15}$  & 12.7  &    10 \\ 
&  outside the BLR & 2.5  &   600  &      8$\times 10^4$   &     2    &   3.6   &    0.3    &     3.2$\times 10^3$    &        1$\times 10^{17}$  &    15  
&      12 \\ 
\hline
\multirow{2}{*}{2011 Jun (two zones)} & internal region & 25  &  610   &     3$\times 10^4$   &      2    &   3.6    &    1.45    &    3.1$\times 10^5$       &  
     1.1$\times 10^{16}$   &  10  &    10 \\ 
& external region &   35   &   610    &    3$\times 10^4$    &    2    &   3.35  &     0.8  &     1.05$\times 10^3$      &         1.5$\times 10^{17}$    &  
 10    &    10 \\ \hline
 \end{tabular}
 \end{table*}

\subsubsection{June 2011: high-activity state}

In June 2011 the multifrequency light curves (Figure~\ref{mw_lc})
show a higher state than during the previous period and significant variability in all bands.
X-ray and HE $\gamma$-rays show a similar behaviour
with two minor flares in May. The optical light curve has only one outburst that happens
during the descending phase of the second peak when both the X-ray and the HE fluxes are decreasing. In the two radio bands there are no clear indications of simultaneous flares with $\gamma$-ray, X-ray, or optical (the gaps present during the $\gamma$-ray and X-ray flares should be noticed). However, there is a fast flare in the 37\,GHz light curve which is simultaneous with the third peak (one bin long) in the HE $\gamma$-ray light curve. In the two radio bands there are gaps during the $\gamma$-ray and X-ray flares. 

A previous study of the multifrequency behaviour of \object{3C~279} \citep{2012Hayashida}
found a correlation between the optical and HE $\gamma$-ray bands
and an absence of correlation between the X-ray and HE $\gamma$-ray
bands between 2008 and 2010. The finding is also in agreement with the tendency of FSRQs to have
correlated emission between optical and HE $\gamma$-ray bands
\citep[e.g.][]{Abdo1502, Abdo1510}. Here a correlation study by selecting data pairs from different light curves with separation of $<0.5$ days yields no significant correlations. Visual inspection (above) suggests a different behaviour of \object{3C~279} in 2011 May and June compared to 2008-2010. In particular it seems there is no correlation between the optical and HE $\gamma$-ray bands. 
 A possible explanation could be a time lag between the emission in the two energy bands
\citep[e.g.][]{Janiak2012}.   
However, this is not likely since the shape of the flux increase
(around MJD 55715) in the HE $\gamma$-ray light curve is different
from the one in the optical R-band. In addition, the optical light curve shows a
quiescent behaviour before the flare, which is difficult to reconcile
with the behaviour observed in the \lat light curve. It should be mentioned that the optical/$\gamma$-ray correlation showed inconsistent patterns as far back as the EGRET era \citep{Hartman2001_corr} and this change of mode (with appearing and disappearing correlations) is, for instance, in agreement with  the long term studies of the source in optical and X-rays \citep{Chatterjee2008}. In summary, the behaviour observed in the light curves in different energy ranges suggests the presence of three different emission regions, one responsible for the radiation in X-ray and HE $\gamma$-ray, a second for the optical, and a third one for the radio emission (see discussion in Sects.4.1.1 and 4.2.).

The co-spatiality of HE and VHE emission and the location inside the BLR of the corresponding emission region is compatible with the Fermi spectrum and MAGIC upper limits. The EBL corrected MAGIC $\gamma$-ray spectrum of the June observations (see Figure~\ref{fermi_magic}) shows that the 95\% confidence upper limits from MAGIC are barely consistent with the 68\% contours of the \lat spectrum, suggesting the presence of a spectral break. The uncertainties in the used EBL model, which are difficult to quantify, are not taken into account. A possible explanation of this feature is that the emission in HE and VHE $\gamma$-rays is generated in the same region, in which a population of low-energy photons is also present. These low-energy photons will interact mostly with the more energetic HE or VHE $\gamma$-ray photons, causing their absorption. 

\begin{figure}
  \centering
\includegraphics[width=88mm]{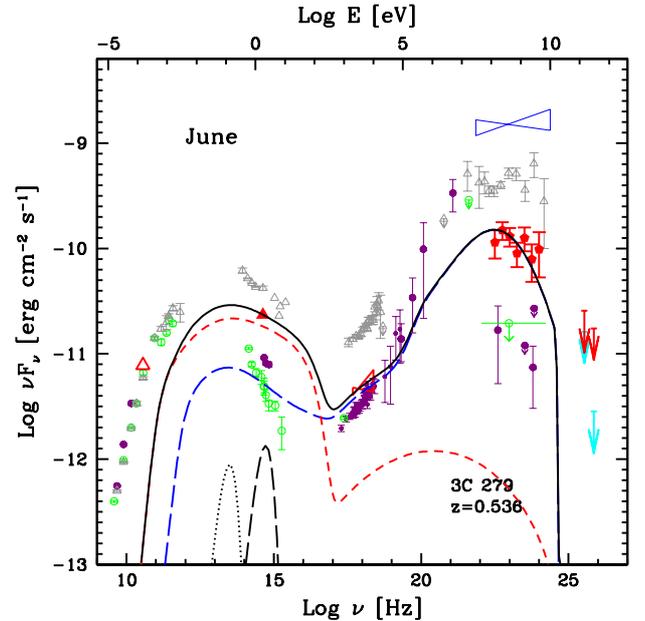}
\caption{June 2011 multifrequency observations. See Figure~\ref{sed_fa} for the data description. The SED is fitted using a two-zone leptonic model. The high-energy emission is dominated by the region inside the BLR (blue long-dashed line) while the synchrotron is dominated by the external region (red short-dashed line) far outside the BLR and the infrared torus. Thus, only the SSC hypothesis is considered for the high-energy bump. The blackbody radiation from the BLR (dashed) and from the infrared torus (dotted) are also shown. The parameters are summarised in Table~\ref{fitsed}. }
\label{two_zones}   
\end{figure}

The features observed in the optical polarisation rotation
suggest an (optical) emission region at distances of about 3\,pc (see Sect.4.2.), but closer to the central engine than the radio core. 
Consequently, we fitted the SED with a two-zone leptonic model
(Figure~\ref{two_zones}): the high-energy emission is dominated by the
region inside the BLR (blue long-dashed line), while the synchrotron radiation,
responsible for the optical and low-energy emission (red short-dashed line), is produced in a
region outside the BLR and infrared torus and therefore  only the SSC scenario 
is considered for the second bump \citep[for details see][]{Aleksic2011a}. However, the size of the external region is
fixed by the variability timescale in the optical to $R=1.5\times
10^{17}\,$cm. Assuming $K=2\times10^3$ and $B=0.5\,$G, values of the order of the ones derived from the SED fit of
the external region, we find that the emission region is opaque to radio frequencies below 100$\,$GHz. Thus, 
the radio data are not included in the fit. The
multifrequency light curves of this epoch do not show evident
correlation between the optical and radio frequencies and therefore
support this scenario.

\subsection{The geometric interpretation of the optical EVPA rotation}

\begin{figure}
  \centering
  \includegraphics[width=88mm]{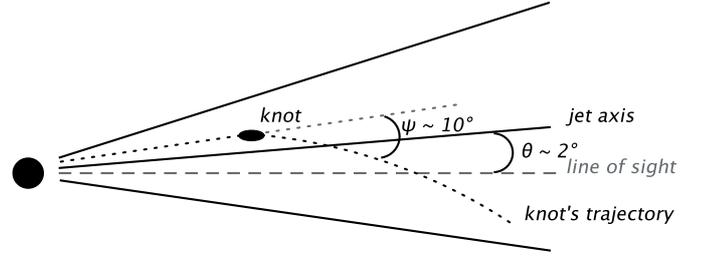} 
\caption{Sketch of the bent trajectory model. An emission knot moving with the flow enters a region where its trajectory bends such that the flow direction crosses the line of sight.}

\label{bent_jet}  
  \end{figure}

\begin{figure}
  \centering
  \includegraphics[height=88mm, angle=270]{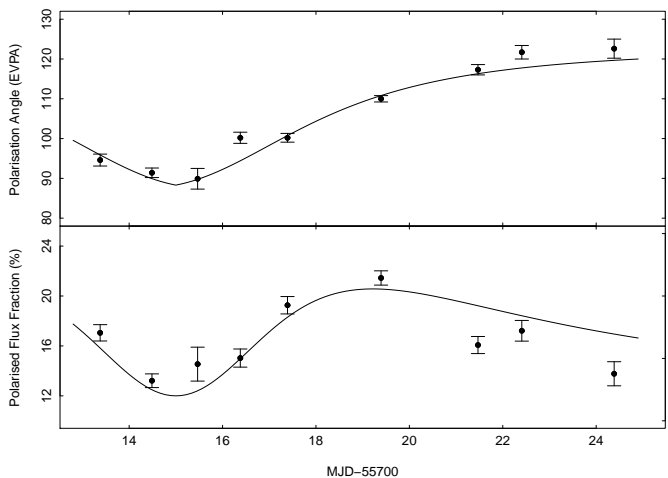}
\caption{Fit of the bent trajectory model to the polarisation quantities at the dates corresponding to the optical flare.
 Both the observed polarisation degree and the percentage of polarised flux closely follow the behaviour predicted 
by the bent trajectory model (Figure~\ref{bent_jet}). The model, as described in the text, is obtained following the theory of \citet{Nalewajko2010}. After the epochs reported in the figure, the monitoring is less intense and the behaviour of the polarisation degree and EVPA seem to change and no longer follow the simple bent trajectory model.}
\label{model_fit}
 \end{figure}

The rotation of $>140^\circ$ of the optical EVPA, which took place simultaneously with the optical outburst and which was accompanied by the increased degree of polarisation between MJD 55710-730, can be explained with purely geometric and relativistic aberration effects (Figure~\ref{bent_jet}). In detail, the EVPA light curve for the 10 days around the optical flare (Figure~\ref{model_fit}) shows a smooth rotation by $\sim15^{\circ}$ in the clockwise direction, followed by an inversion and a smooth $40^{\circ}$ EVPA change in the counter-clockwise direction. 

We propose a scenario (Figure~\ref{bent_jet}) in which an emission knot moving with the flow enters a region where its trajectory bends. Since the jet is closely aligned to the line of sight (l.o.s.) by about $2^{\circ}$ and the plasma flows relativistically, with $\Gamma \lesssim 16$ \citep{Wehrle2001, Jorstad2005}, small bends $\Psi$ of only a few degrees are enough to produce strong effects both in the apparent deflected angle and in the observed flux and polarisation because of relativistic aberration. In particular, for this range of speeds, a bend of $\Psi \lesssim 10^{\circ}$, from $-2^\circ$ to $8^\circ$, will imply a total \textit{apparent} bending as seen in the projected EVPA direction of $\sim60^{\circ}$. Furthermore, if the bend is in a direction such that the trajectory axis crosses the l.o.s., two main effects will follow. First, after the trajectory crosses the l.o.s., the observer will perceive the EVPA rotating in the opposite direction, with the profile shown in Figure~\ref{model_fit}. The inversion simply results from projection effects, and the specific profile depends on the speed of the flow. Secondly, once the trajectory axis approaches the l.o.s. direction, the polarisation degree is expected to decrease because a slight angle between the axis and the l.o.s. favours an apparent symmetry of the field as it is projected on the plane of the sky. After the polarisation degree passes through a minimum coinciding with the crossing of the l.o.s., it increases again, until the observer leaves the radiation cone of the jet. After the bending exceeds $1/\Gamma$ relative to the l.o.s., the polarisation decreases, its maximum being attained when the l.o.s. crosses the border of the radiation cone. In this model, we used $\Gamma \sim 16$, corresponding to a half-opening angle of the radiation cone $\phi \sim 4^{\circ}$. 

As shown in Figure~\ref{model_fit}, the observed polarisation degree closely follows this behaviour. The solid model lines result from the comparison of the data and the predicted model built from the characteristics of the EVPA rotation and the jet parameters as measured by VLBA and quoted above, not from a minimisation fit as it would be difficult because of the degeneracy of many of the parameters involved. However, once  all values derived from VLBA and MWL observations are put in, the only free parameter left to try to reproduce the behaviour of the polarisation degree is the ratio of transversal to axial magnetic field components, $b$ \citep[see][]{Barres2010}, which we found to imply a dominance of the axial field component, $b < 0.5$, in order to reproduce the observed behaviour. 

In this scenario we also expect an increase of the photometric flux as the trajectory of the emitting region crosses the l.o.s. The fractional change in the observed (aberrated) photometric flux is given by \citet{Urry1995} $F_{los}/F_0 \sim (\delta/\delta_0)^{(3+\alpha)}$, where $\alpha$ is the radio spectral index defined as $F_\nu \sim \nu^{-\alpha}$ (we use the exponent $3+\alpha$ because we are dealing with fluxes in a narrow energy band, and hence quasi-monochromatic), the Doppler factor $\delta = [\Gamma(1-\beta~cos\theta)]^{-1}$, is a function of the viewing angle $\theta$. Using the parameters for the jet derived in Table~\ref{fitsed} for the external region (June), we estimate that the change in flux between the initial position ($\theta \sim 2^{\circ}$) and the flux at the site when the trajectory crosses the l.o.s. due to the bend is $F_{los}/F_{0} \approx  1.1 - 1.2$, for a spectral index $\alpha = -2.5$. It can be readily seen that the flux change induced by the aberration effect is too small to explain the nearly doubling of the optical flux registered in the light curve. This is not in contradiction to the neat fit obtained with the aberration model to the behaviour of the polarisation quantities, but it suggests that more than one process (such as an increase in the total synchrotron emissivity of the region without significant changes in $B$) is taking place simultaneously with the change in the trajectory. 

Finally, we can apply this model to put limits on the location of the event. The total bending angle is $\Psi \approx 10^{\circ}$ and the rotation lasts 14 days. This implies a rotation rate of $\approx 0.9^\circ$ per day, corresponding to the $\approx 4^\circ$ per day of apparent EVPA rotation. For $\Gamma = 16$, the linear distance travelled by the blob during the event can be estimated to be $\Delta r \sim c\Delta t \Gamma^2 \sim   9.25\times10^{18}$~cm -- or $\sim3$ pc. Since the path of the blob is curved, the de-projected linear distance travelled by the blob can be related to the position of the bend by $r_{\rm{bend}}  \gtrsim \Delta r$ \citep{Nalewajko2010}. 

There are different possible explanations for the bent trajectories of the moving emission feature \citep{Abdo2010Nature}: a physical bend in the jet  caused by, for example, hydrodynamical ram pressure equilibrium with the outer medium \citep{Hardee1990}, or flow through and helical magnetic field of the jet \citep{Marscher2008,Larionov2008}. For single rotation it is not possible to identify the origin of the bent trajectory. However, the detection of earlier gradual and smooth rotation of the EVPA during high-activity states \citep[][at opposite directions]{Larionov2008,Abdo2010Nature} 
led \citet{Abdo2010Nature} and \citet{Nalewajko2010} to propose that there might be a bend in the jet of \object{3C~279} that is responsible for the observed change in the sense of the EVPA rotation. The bend itself would be located somewhere between the emission region observed by the two groups. The longer timescales associated with the rotation observed by \citet{Larionov2008} suggest that their event is located farther out in the jet (lower limit $\sim20\,$pc), whereas the event observed by the \lat was estimated to happen within the first $10^{19}$ cm, corresponding to a scale of $\approx 3$ pc, and containing the blazar zone ($R_{\rm{Blazar}} < 10^{17}$~cm) associated with the size of the GeV $\gamma$-ray emitting region as observed at the time. 
Adapting this interpretation to our results suggests that the bend must happen in the first 3 pc. 
Using the relation $R_g = 2GM_{SMBH}/c^2\sim 3\times 10^{11-12}$, with the mass of the supermassive black hole $M_{SMBH} \sim 10^{8-9}\, M_{\odot}$ \citep{Nilsson2009}, the location of the bend can be expressed in units of gravitational radii: 
$\sim3 \times 10^{4-5} R_{g}$. The radio core is estimated to be $\sim10^5 R_{g}$ \citep{Marscher2008} consequently,
the bend is positioned near the radio core.

\section{Summary and conclusions}

The flat spectrum radio quasar \object{3C~279} was observed in 2011 with the MAGIC telescopes in two different campaigns: first with regular monitoring from February to April and later in June as follow up observations after high-activity states in optical and HE $\gamma$-ray bands. In neither of the periods  was a detection found; consequently, differential upper limits were calculated.  

Simultaneously with the MAGIC observations, \object{3C~279} was monitored at lower energies: HE $\gamma$-rays, X-ray, optical, and radio. We examine the multifrequency light curve obtained merging all the available information. There are various periods 
of enhanced activity in all the examined energy ranges. 
The HE $\gamma$-ray and X-ray time evolution show similar behaviours, with two flares both occurring before the optical outburst. Simultaneous with the outburst, an increase in the optical polarised flux and a smooth rotation of the EVPA was observed. 
\begin{large}\end{large}
We compiled broadband SEDs for the two different MAGIC observations periods, and fitted them with leptonic models taking into account the variability patterns in the multifrequency data. For the low state of the source we cannot constrain the location of the emission region, neither from the light curves nor from the SED modelling.
For the period of the higher activity, the similar trend in \lat and X-ray light curves suggests that the emission observed in these two energy ranges originates from the same region, different from that where the optical emission takes place. In addition, the indication of a cutoff in the GeV range hints that the $\gamma$-ray emission is coming from an inner region of the blazar, and therefore internally absorbed in the MAGIC energy range. 
In this context, we fitted the June 2011 data with a two-zone leptonic model. The high-energy emission (from X-ray to VHE $\gamma$-rays) originates from an inner region of the jet, while we locate the optical flare at the parsec scale. The location of the optical emission region is derived from the rate of the rotation of the EVPA of the optical polarisation simultaneous with the optical flare.

We also investigate the feasibility of the geometric interpretation for the observed rotation of the EVPA. We find that the observed rotation is in good agreement with a model where the emission feature follows a bent trajectory. It has been suggested by \citet{Abdo2010Nature} and \citet{Nalewajko2010} that there might be a bend in the jet of \object{3C~279} and we find that our polarisation observations are in agreement with their hypothesis. However, turbulence would cause rotations in opposite directions and cannot be excluded as a cause of the observed rotation. We also note that we do not see a signature of the bend in the observations in other wavelengths, but the signature there could be damped by the other dominating emission regions. This is in agreement with the multiple emission regions that we suggest based on the variability patterns in the other wavelengths (see above). 

The MAGIC observations presented in this paper have provided the strongest upper limits on the VHE $\gamma$-ray emission from the source so far. The upper limits are for the first time below the detected fluxes, confirming that the detections represented a high-activity state of the source in VHE $\gamma$-rays. The source was previously detected at VHE in 2006 and 2007, and both detections were during high-activity states in the optical and X-ray bands. The multiwavelength behaviour observed in June 2011 can be compared with those measured during the observations of 2006 and 2007. In February 2006 (first VHE $\gamma$-ray detection), the optical flux was similar to the one measured in June 2011, while the X-ray flux was lower than in June 2011. No optical polarisation data are available for the VHE detection in 2006. Compared to January 2007, the optical flux in June 2011 was lower, but the X-ray level was similar. In 2007, rotations of the optical EVPA and 43\,GHz VLBA core were detected, while in June 2011 the optical EVPA rotation was not accompanied by the rotation of the EVPA in the radio band. There are some similarities regarding the multifrequency behaviour during the June 2011 flare with respect to that previously observed during VHE detections. Despite these similarities, the recent VHE observations did not yield a significant signal. Given the known fast $\gamma$-ray variability,  the non-detection of VHE emission could be a result of the unfortunate timing of the MAGIC observations (limited by the moon conditions) during the decay phase of the flare detected by \textit{Fermi}--LAT. But it might also mean that VHE emission is rare in \object{3C~279} (even during flares). Resolving the open question of the origin of the VHE $\gamma$-ray emission in \object{3C~279} and flat spectrum radio quasars in general requires long-term observational effort in VHE $\gamma$-ray band.    

\begin{acknowledgements}

We would like to thank the Instituto de Astrof\'{\i}sica de
Canarias for the excellent working conditions at the
Observatorio del Roque de los Muchachos in La Palma.
The support of the German BMBF and MPG, the Italian INFN, 
the Swiss National Fund SNF, and the Spanish MICINN is 
gratefully acknowledged. This work was also supported by the CPAN CSD2007-00042 and MultiDark
CSD2009-00064 projects of the Spanish Consolider-Ingenio 2010
programme, by grant 127740 of 
the Academy of Finland,
by the DFG Cluster of Excellence ``Origin and Structure of the 
Universe'', by the Croatian Science Foundation Project 09/176, by the University of Rijeka Project 13.12.1.3.02, 
by the DFG Collaborative Research Centers SFB823/C4 and SFB876/C3,
and by the Polish MNiSzW grant 745/N-HESS-MAGIC/2010/0.

The \textit{Fermi} LAT Collaboration acknowledges generous ongoing support
from a number of agencies and institutes that have supported both the
development and the operation of the LAT as well as scientific data analysis.
These include the National Aeronautics and Space Administration and the
Department of Energy in the United States, the Commissariat \`a l'Energie Atomique
and the Centre National de la Recherche Scientifique / Institut National de Physique
Nucl\'eaire et de Physique des Particules in France, the Agenzia Spaziale Italiana
and the Istituto Nazionale di Fisica Nucleare in Italy, the Ministry of Education,
Culture, Sports, Science and Technology (MEXT), High Energy Accelerator Research
Organization (KEK) and Japan Aerospace Exploration Agency (JAXA) in Japan, and
the K.~A.~Wallenberg Foundation, the Swedish Research Council and the
Swedish National Space Board in Sweden.

Additional support for science analysis during the operations phase is gratefully
acknowledged from the Istituto Nazionale di Astrofisica in Italy and the Centre National d'\'Etudes Spatiales in France.

The OVRO 40-m monitoring program is supported in part by NASA grants NNX08AW31G and NNX11A043G, and NSF grants 
AST-0808050 and AST-1109911.

The Mets\"ahovi team acknowledges the support from the Academy of Finland
to our observing projects (numbers 212656, 210338, 121148, and others).

\end{acknowledgements}

\bibliography{bibliography.bib}
\bibliographystyle{aa}

\end{document}